\documentclass[%
reprint,
 amsmath,amssymb,
 aps,showkeys,
pre
]{revtex4-1}

\usepackage{graphics}
\usepackage{graphicx}% Include figure files
\usepackage{dcolumn}% Align table columns on decimal point
\usepackage{bm}% bold math
\usepackage{hyperref}% add hypertext capabilities
\usepackage{placeins}

\newcommand\erf{\mbox{erf}}

\begin{document}
%\title{Restart could optimize splitting probabilities of first-passage process}
\title{Restart could optimize the probability of success in a Bernoulli trial}
\date\today

\author{Sergey Belan}
\email{belan@itp.ac.ru}
\affiliation{Moscow Institute of Physics and Technology, 141700 Dolgoprudny, Russia}
\affiliation{Landau Institute for Theoretical Physics of the RAS, 
142432  Chernogolovka, Russia}

 \begin{abstract}
Recently noticed ability of restart to reduce the expected completion time of first-passage processes allows appealing opportunities for performance improvement in a variety of settings. 
However, complex stochastic processes often exhibit 
several possible scenarios of completion 
which are not equally desirable in terms of  efficiency.
Here we show that restart may have profound consequences on the splitting probabilities of a Bernoulli-like first-passage process, i.e. of a process which can end with one of two outcomes. 
Particularly intriguing in this respect is the class of problems where a carefully adjusted restart mechanism maximizes  probability that the process will complete in a desired way.  
We reveal the universal aspects of this kind of optimal behaviour by applying the general approach recently proposed for the problem of first-passage under restart.
 \end{abstract}

% \ocis{240.6680, 160.3918, 260.3910}

\maketitle

Stochastic processes subject to restart appear in many disciplines including physics, chemistry, biology and computer science.
Restart means sudden interruption of a process followed by its start anew. 
In some contexts restart is an integral part of a phenomenon under study (e.g. substrate unbinding in enzymatic reactions \cite{Reuveni_2014} and  recovery of RNA polymerase from the backtracked state \cite{Roldan_PRE_2017}), while in other  it plays a role of an external control tool (e.g. reinitialization of a randomized computer algorithm \cite{Luby_1993,Montanari_2002} and reduction of growing tumor  to its initial size by chemical treatment \cite{Gupta_2014}).  

Significant amount of research effort has been dedicated towards study of the effect of restart on the first-passage properties.
The growth of interest in this problem was triggered by the surprising observation that restart may significantly reduce the mean first-passage time (MFPT).
Over recent years it has been demonstrated in a range of diverse examples that a carefully chosen restart rate can bring the MFPT to a minimum \cite{EM_2011,Evans_2011,Whitehouse_2013,Evans_2014,Kusmierz_2014,Kusmierz_2015,Pal_2016,Eule_2016,Nagar_2016}.
Along with the investigation of particular cases, we witness ongoing attempts to reveal the general principles allowing to navigate in a vast space of  first-passage problems under restart.  
%Along with the investigation of particular cases, there are ongoing attempts to develop an unified way of thinking about entire space of first passage problems under restart. 
Remarkable result of those attempts is the discovery of universality displayed by all optimally restarted processes \cite{Reuveni_PRL_2016,Reuveni_PRL_2017}.

To the best of our knowledge, first-passage processes under restart considered so far had only one way of completion.
%To the best of our knowledge, all previous studies of first passage under restart consider single-outcome processes, i.e.  processes having only one way of completion. 
Say, diffusion mediated search with stochastic resetting to initial position \cite{EM_2011} -- a classical example of a  first-passage problem under restart --  ends if and only if a searcher finds a target.
%However, in many real-life settings there are variety of ways in which stochastic process can end. 
However, real-life settings often offer a variety of possible ways in which stochastic process can complete.
Plurality of the process outcomes may arise from the competition among several different first-passage  phenomena or due to multiple thresholds for one and the same first-passage mechanism.
Assume, for instance, that gambler stops playing after winning a certain amount of money or getting ruined, whichever happens first \cite{Edwards_1983, Feller_1968}. 
In many-target search problems and diffusion-limited reactions, different completion scenarios may correspond to finding of different targets \cite{Condamin_2007, Condamin_2007_1, Condamin_2008,Meyer_2012,Calandre_2012,Benichou_2015,Calandre_2014}.
% one could ask which of the targets is reached first ////one may distinguish the 
%Biological species cease to exist  either through  extinction or when speciation events occur.
In search problems with time constraints, a search process can finish either by target detection or by searcher/target death \cite{Abad_2012,Yuste_2013, Abad_2013,Abad_2013_1,Abad_2015,Campos_2015,Meerson_2015,Grebenkov_2016_1}.
When there are several competitive paths of chemical reaction, an individual molecule may be converted into one product or other depending on which path has been realized \cite{Rehbein_2011,Rehbein_2015,Martin-Soomer_2016}.
%Similarly, a protein may fold into one of multiple distinct native states.
Similarly, a protein may fold along one of many possible pathways to one of multiple native states \cite{Solomatin_2010,Marek_2011,Hyeon_2012,Paudel_2014,Hinczewski_2016,Pierse_2017}.
In evolutionary biology and ecology, one could ask if a population goes extinct before its size attains some threshold level   \cite{Lande_2003,Drake_2009}.
% (carrying capacity)
%The existence of the species may end as a result of extinction or through speciation \cite{Newman_2003,Chowdhury_2004,Yamaguchi_2013,Hagen_2015}.
%Existence of biological species may terminate  either through  extinction or speciation.
%Biological species cease to exist either through extinction or speciation \cite{Newman_2003,Chowdhury_2004,Yamaguchi_2013,Hagen_2015}. 
%Different ending scenarios may correspond, for instance, to detection of different targets in many-target search problem.
%Similarly, the search process with time constraints may finish either by target detection or by searcher/target death.
%An individual enzyme molecule in a enzymatic reaction with inhibition may either form a complex or meet the inhibitor molecule.
%Another important possibility is that different outcomes may correspond to different products of chemical reaction. 
Clearly, the immense set of possibilities is not limited to these few examples.

What happens when a first-passage process with several possible outcomes becomes subject to restart?
%The probability of observing a particular completion scenario of a first-passage process is known as  the  splitting probability.
%Beside first-passage time, multi-outcome processes are characterized by the so-called splitting probabilities which are defined as the probabilities of getting specific
%outcomes 
%before getting any of other outcome 
%from the full spectrum of alternatives.
The main goal of this Letter is to draw attention to previously unknown type of optimal behaviour in first-passage phenomena: a carefully chosen rate of Poisson restart brings the probability of observing a particular completion scenario to a maximum (or minimum).
In other words, we argue that stochastic restart could optimize the so-called \textit{splitting probabilities} \cite{Kampen_1992,Redner_2001}.
The effect is first illustrated on a particular example and after that 
we apply a general framework recently proposed in Ref. \cite{Reuveni_PRL_2017} by Pal and Reuveni  to gain a more deep insight. 
%The general approach helps us to %formulate the general conditions of existence of optimal restart rate and 
%uncover universal properties shared by all optimally restarted processes irrespective on their fine details.
For the sake of simplicity we focus on the case where the process has exactly two possible outcomes, but the analysis can directly be extended to a more general situation. 
We show that optimality of the splitting probabilities always entails an exact match between the unconditional and conditional mean completion times of the process.  
%We show that when the restart rate is optimal, the  MFPT conditional to the outcome being optimized is always equal to the unconditional MFPT of the process.   
Looking for further generalization, we go beyond the assumption of Poisson restart and demonstrate advantage of  the deterministic restart strategy  in terms of attaining the most pronounced extrema of splitting probabilities.

The key properties of first passage under restart have been originally learned from the one dimensional diffusion process \cite{EM_2011}.
We will use the same  "Drosophila" to demonstrate the ability of stochastic restart to optimize the splitting probabilities.  
Specifically, let us consider a mortal Brownian searcher with the diffusion constant $D$ and the mortality rate $\alpha$  which starts from the initial position $x_0\ge 0$.
The search process ends when either searcher dies or when it finds a target located at $x=0$.
It is shown in Ref. \cite{Abad_2013_1} (see also \cite{Meerson_2015}) that target detection occurs with the probability $p=e^{-\sqrt{\alpha x_0^2/D}}$.
%Exploiting results of \cite{Roldan_PRE_2017}, where mathematically equivalent problem has been examined, we find that target detection occurs with the probability $p=e^{-\sqrt{\alpha x_0^2/D}}$.
%Expectedly, $p$ approaches  unity as $\alpha\to 0$ and vanishes when $\alpha\to\infty$. 
Assume now that the  process is stochastically restarted, i.e. the searcher is returned to its initial position $x_0$ at some constant rate $r$ \cite{note_1}. 
What is the detection probability $p_r$ in the presence of restart?
%How does the detection probability is modified in the presence of restart?
Exact solution of the initial-boundary value problem for the probability density of the searcher' position yields (see Supplemental Material)
% the following expression for  we find that detection probability  that searcher will find target is given by 
%the probability $p_r$ that the restarted search process will end with target detection is given by 
% the probability that the target will be detected  (see Supplemental Material)
\begin{equation}
\label{p_r}
p_r=\frac{r+\alpha}{\alpha e^{\sqrt{(r+\alpha)/D}x_0}+r}.
\end{equation}
%Correspondingly, the probability that searcher dies is $q=1-p$.
Analysing Eq. (\ref{p_r}),
%the behaviour of $p_r$ in dependence on the restart rate $r$, 
one can readily see that if $\alpha\ge \alpha_0=(z^*)^2D/x_0^2$, where $z^*\approx 1.59362...$ is the solution to $z/2=1-e^{-z}$, then $p_r$ monotonically decreases as $r$ increases from zero to infinity.
Otherwise, when $\alpha< \alpha_0$, the probability $p_r$ takes its maximum at the non-vanishing restart rate $r_0=\alpha_0-\alpha$.
In Figure \ref{p_r} we plot $p_r$ as a function of $r/\alpha_0$ for different $\alpha/\alpha_0$.

\begin{figure}
 \includegraphics[scale=.217]{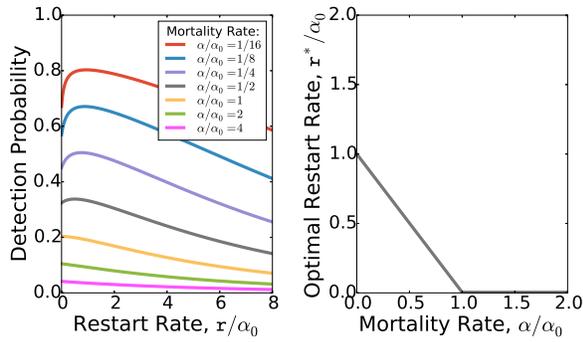}
 \caption{The probability of target detection $p_r$ as a function of the rate $r$ of Poisson restart for different values of the decay constant $\alpha$ of the mortal searcher.}
  \label{pic:p_r}
 \end{figure}

Let us give a qualitative explanation for the observed behaviour of $p_r$.
If  $\alpha$ is large compared to $D/x_0^2$, then the typical size of the region explored by the searcher during its lifespan is less than the initial distance to the target and, thus, a non-vanishing restart inevitably leads to reduction of the search efficiency.
Otherwise, when $\alpha$ is small in comparison with $D/x_0^2$, the searcher leave long enough to be able to reach target via typical diffusive path, but it is also able to execute distant excursion in empty areas of the search space.
These excursions prolong the search process and typically end with searcher death. 
Then, the non-vanishing restart rate censors the fatal paths and increases  chances to find the target.
On the other hand, too large restart rate hinders target detection since the searcher has less time between restarts to reach the origin under the same mortality rate. 
This is why there exists a non-vanishing optimal restart rate $r^*$ which brings the probability  that searcher will find the target before dying to a maximum.

%Restart time scale $r^{-1}$ plays the role of an exponential upper cutoff of stochastic process time: the contribution of realizations of stochastic process with completion times that are much longer than the typical restart time is effectively censored out.
%At $\alpha<r^*$ the prolonged realizations of the stochastic process preferentially lead to the undesirable outcome, i.e. particle die when it go far away from the wall.
%In this case, non-vanishing restart rate censors these realization and increases the probability to find target.
%. 
%Below we formulate a general condition of existence 
%
%Before we pass to the general treatment, let us consider the first-passage time. 
%Each outcome is characterized by the conditional first-passage time.
%

%Therefore, the optimal resetting rate is given by  
%\begin{equation}
%r^*=\left\{ \begin{array}{ll}
%0,\ \ \ \ \ \ \ \ \   \text{if}\ \ \ \  \alpha> r^*\\
%r^*-\alpha,  \ \ \text{if}\ \ \ \   \alpha< r^*
%\end{array} \right.
%\end{equation}
%where $r^*=(z^*)^2D/L^2$ is the Evans-Majumdar optimal rate ($z^*=1.59362...$ is the solution of equation $z/2=1-e^{-z}$) \cite{EM_2011}.

\begin{figure}
 \includegraphics[scale=0.4]{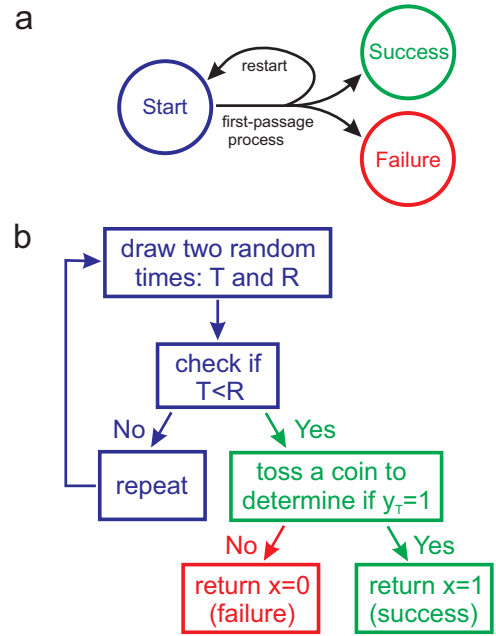}
 \caption{\textbf{a:} Bernoulli-like first-passage process under restart \textbf{b:} Pseudocode representation of  Eq (\ref{eq_x}).}
  \label{pic:scheme}
 \end{figure}

%\textbf{Splitting probabilities of a generic process under restart.}
Having examined %the main idea on 
 the exemplary case, we now turn to more general setting.
Let us consider a generic stochastic process which can end in two incompatible ways and is subject to a generic restart mechanism.
For the sake of convenience we will call one of two possible outcomes as success and the other one as failure.
Thus, the problem can be viewed as a kind of Bernoulli experiment, see Fig. \ref{pic:scheme}a.
In the example discussed above, detection of target naturally corresponds to success, while searcher' death is interpreted as failure.
Obviously, in other contexts these conventional terms may not have any real meaning.

%The process is subject to restart mechanism which make process to start from the stretch. 
%, so that the whole setting can be schematically . 
%The setting is schematically illustrate in Fig. 1a. 

%The analytical approach utilized below is applicable to a generically distributed restart time $R$.
%but we first consider the case of exponential distribution $re^{-rR}$.
%By other words, the restart events have Poisson statistics with rate parameter $r$.
%Then the restart time $R$ has exponential distribution $re^{-rR}$. 
%More general setting is discussed at the end of the Letter. 

%Now let us assume that the process is subject to stochastic restart. 
%The random restart time $R$ is exponentially distributed with rate
%parameter $r$.
%If $T<R$, then the process is completed prior to restart.
%Otherwise, the process will start from beginning until it reaches completion.

The original  process is characterized by a random completion time $T$ having the probability distribution  $P(T)$.
%Incompatibility of different ending scenarios allows us represent
The later can be decomposed into a sum $P(T)=P^s(T)+P^f(T)$, where $P^s(T)$ and $P^f(T)$ are the probability densities of successful and failed trials, respectively. 
%In particular, for the above example we have  and , see Fig 3. 
Normalization of the probability density $P^s(T)$ defines the "unperturbed" probability $p$ of success:
%successful process completion  
$p=\int_0^\infty P^s(T)dT$.
Conservation of total probability implies that $\int_0^\infty P^f(T)dT=1-p$.
We will also utilize the trivial fact that the ratio $P^s(T)/P(T)$ gives the probability of success in a trial with the completion time $T$.   %success for the process realization   with completion time $T$.
%The normalization conditions for these functions are $\int_0^\infty P(T)dT=1$, $\int_0^\infty P^s(T)dT=p$ and $\int_0^\infty P^f(T)dT=q$,
%where $p$ and $q=1-p$ are the splitting probabilities.

Being subject to restart, the process can be interrupted at a random time $R$, characterized by a proper probability distribution $P^r(R)$, and started again. 
%The probability distribution of the restart time $R$ is denoted as $P^r(R)$.
The probability $p_r$ of success for the restarted process can be computed as expectation of a binary random variable $x$ which takes the value $1$ if the process is completed in success and is equal to $0$ in the case of failure. 
%that the outcome of the restarted process will be successful.
% Otherwise, the process will start
%from scratch and begin completely anew
% will start from stratch until it reaches completion.
%It is convenient to introduce a Bernoulli random variable $x$ which takes the value $1$ if process end by  success and is equal to $0$ in the case of failure.
This variable obeys the following renewal equation:
\begin{equation}
\label{eq_x}
x=I(T<R)y_T+I(T\ge R)x',
\end{equation}
where $I(T<R)=1-I(T\ge R)$ is an indicator random variable which is equal to unity when $T<R$ and
is zero otherwise; $x'$ is an independent and identically distributed
copy of $x$; $y_T$ is an auxiliary binary variable which takes the value one with probability $P^s(T)/P(T)$.

The intuition behind Eq. (\ref{eq_x}) is very simple. 
Imagine that we run a computer simulation designed to reproduce  behaviour of the random variable $x$. 
At the first step, we should choose two random times from the distributions $P(T)$ and $P^r(R)$ and decide
which of the two, restart or completion, happened first.
If $T<R$, then the process is completed prior to restart.
To determine whether the process end in success or in failure we toss a coin with probability of success $P^s(T)/P(T)$ and assign the outcome to the variable $x$.
Otherwise, if $T<R$, the process begins completely anew and we should repeat the procedure until the process reaches completion.
This scheme is best illustrated in the form of pseudocode, see Fig. \ref{pic:scheme}b.

%We proceed further by noting that the probability of success in a Bernoulli trial is just the expected value of a Bernoulli variable, i.e. $p_r=\langle x\rangle$.
%Here and in what follows, angular brackets denote averaging over the statistics of the underlying process and random restart events. 
After averaging  over the statistics of the underlying process and random restart events,  Eq. (\ref{eq_x}) yields  
\begin{equation}
\label{mean_x}
p_r=\langle x\rangle=\frac{\langle I(T<R)y_T\rangle}{\langle I(T<R)\rangle}.
\end{equation}
Once the probability density functions $P^s(T)$ and $P^r(R)$ are known, one can readily compute $\langle I(T<R)y_T\rangle=\int_0^\infty\int_0^R  P^r(R)P^s(T)dRdT$ and $\langle I(T<R)\rangle=\int_0^\infty\int_0^R  P^r(R)P(T)dRdT$.
%Thus, Eq. (\ref{mean_x}) allows us also to find the probability of success of an arbitrary stochastic process under a generic restart mechanism.
%Obviously enough, Eq. (\ref{mean_x}) is not modified in the presence of a generally distributed random time penalty for restart (See Supplementary Material).
%Let us look at the case 
When restart events posses Poisson statistics with constant rate parameter $r$,  the restart time $R$ has exponential distribution $P^r(R)=re^{-rR}$ and 
%In other words, .
%Equation (\ref{mean_x}) allows one to explicitly compute the probability of success once statistics of restart events and underlying process are know.   
%Since restart time $R$ has exponential distribution $re^{-rR}$  by assumption, 
%It is straight forward to show then that 
Eq. (\ref{mean_x}) reduces to 
\begin{equation}
\label{splitting_probability}
p_r=\frac{\tilde P^s(r)}{\tilde P(r)},
\end{equation}
where $\tilde P^s(r)$ and $\tilde P(r)$ denote the Laplace transforms of, respectively,  $P^s(T)$ and $P(T)$ evaluated at $r$.
Note that for the above problem of diffusion mediated search $P^s(T)=\sqrt{x_0^2/4\pi DT^3}e^{-\alpha T-x_0^2/4DT}$ and $P^f(T)=\alpha e^{-\alpha T}\erf(\sqrt{x_0^2/4DT})$ \cite{Roldan_PRE_2017}. % where $\erf$ denotes the error function.
It is straight forward to show then that Eq. (\ref{splitting_probability}) reproduces Eq. (\ref{p_r}) previously obtained through the less generic method (see Supplementary Material).  

\begin{figure}
 \includegraphics[scale=0.206]{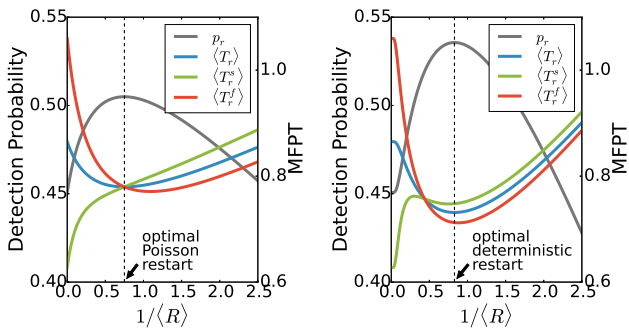}
 \caption{The probability of target detection $p_r$ and the MFPTs, $\langle T_r\rangle$, $\langle T_r^s\rangle$ and $\langle T_r^f\rangle$, versus the inverse mean restart time $1/\langle R\rangle$ for the mortal Brownian searcher under Poisson restart (left) and  deterministic restart (right).  Here we take $D=1$, $x_0=1$ and $\alpha=\alpha_0/4$. 
Interestingly, in both cases $p_r$ takes its maximum value when $\langle T_r\rangle$ attains minimum.
This observation is explained by the relation $p_r=1-\alpha \langle T_r\rangle$ which is valid whenever the mortality rate $\alpha$ is time-independent.}
  \label{pic:MFPT}
 \end{figure}

We are mostly interested in the class of problems where probability $p_r$ is maximised at a nonvanishing optimal rate $r^*$ of Poisson restart.
In principle, one can construct an infinite number of examples belonging to this class.
%In principle, there is a countless number of first-passage processes belonging this class. 
%In principle, this class includes a countless number of first-passage processes.
What do they all have in common?
To address this question let us take a look at the first-passage-time properties of the process illustrated in Fig. \ref{pic:scheme}a. %two-outcome process subject to restart.
%Equivalently one can say that probability of failure is minimized.
%What can we say about dynamics of a two-outcome process under restart?
%To  let us take a look at the dynamics of a two-outcome process under restart.
As it is shown in \cite{Reuveni_PRL_2017}, 
%A general treatment the first passage under restart was given in \cite{Reuveni_PRL_2016,Reuveni_PRL_2017}. 
the completion time $T_r$ of a generic first-passage process under a generic restart mechanism obeys the following identity 
\begin{equation}
\label{fpt}
T_r=I(T\ge R)(R+T_r')+I(T<R)T,
\end{equation}
in which $T_r'$ is an independent and identically distributed
copy of $T_r$. 
Equation (\ref{fpt}) allows one to express the MFPT as $\langle T_r\rangle=\langle\min(T,R)\rangle/\langle I(T<R)\rangle$,  where $\min(T,R)$ is the minimum of $T$ and $R$.
%In the model with exponentially distributed restart time, the mean first passage is found to be 
%\begin{equation}
%\label{mfpt}
%\langle T_r\rangle=\frac1r\frac{1-\tilde P(r)}{\tilde P(r)}.
%\end{equation}
Next, one could ask also how to compute the MFPT  $\langle T_{r}^s\rangle$ conditional to success, which is simply the average completion time of successful trials.
By virtue of its definition, this quantity can be written as  $\langle T_{r}^{s}\rangle={\langle xT_r\rangle}/{\langle x\rangle}$.
Substituting  Eqs. \ref{eq_x} and (\ref{fpt}) into this relation results in
\begin{equation}
\label{MFPT}
\langle T_{r}^s\rangle=\frac{\langle I(T>R)R\rangle}{\langle I(T<R)\rangle}+\frac{\langle I(T<R)y_T T\rangle}{\langle I(T<R)y_T\rangle}.
\end{equation}
For exponentially distributed restart, Eq. (\ref{MFPT}) takes a particularly simple form (see Supplementary Material)
\begin{equation}
\label{conditional_MFPT}
\langle T_{r}^s\rangle=\langle T_r\rangle-\frac{d \ln p_r}{dr},
\end{equation}
where $p_r$ is given by Eq. (\ref{splitting_probability}) and $\langle T_r\rangle=r^{-1}(1-\tilde P(r))/\tilde P(r)$ \cite{Reuveni_PRL_2016}.  
If the success probability  $p_r$ of the restarted process attains a  maximum at some $r^*$, the second term in the right hand side of Eq. (\ref{conditional_MFPT}) vanishes and we get 
%we have $[d\ln p_r/dr]_{r^*}=0$ % [d\ln (1-p_r)/dr]_{r^*}=
 %and therefore 
\begin{equation}
\label{MFPT_identity}
\langle T_{r^*}^s\rangle=\langle T_{r^*}\rangle.
\end{equation}
 Also, since $p_r\langle T_{r}^{s}\rangle+(1-p_r)\langle T_{r}^{f}\rangle=\langle T_r\rangle$, similar identity holds true for the mean completion time of failed trials: $\langle T_{r^*}^f\rangle=\langle T_{r^*}\rangle$. %$\langle T_r\rangle-d\ln (1-p_r)/dr$.
We thus conclude that when the rate of Poisson restart is optimal, in the sense that it maximizes or minimizes the probability to observe specific outcome, the unconditional MFPT is equal to the MFPT conditional to this outcome. 
This universal feature is shared by all optimally restarted processes irrespective on their fine details (see the left panel of Fig. \ref{pic:MFPT} for illustration).
%This conclusion is universal, i.e. it is valid for any optimally restarted process independently on its fine details.
%Moreover, Eq. (\ref{MFPT_identity}) holds true in the presence of generally distributed random time penalty for restart, see SI.   

Surprising simplicity of Eq. (\ref{MFPT_identity}) calls for its intuitive explanation.
To provide such an explanation let us assume that one starts to observe a first-passage process, which is allowed to repeat itself over and over, at a random moment of time. 
What is then the \textit{expected probability} $p^{\text{exp}}$ of getting success in the next outcome?  
It can be shown that this probability is given by $p^{\text{exp}}=p \langle T^s\rangle/\langle T\rangle$ (see Supplementary Material).
Obviously, applying Poisson restart with infinitesimally small rate $\delta r$ will increase the chances of success whenever $p^{\text{exp}}<p$, while at $p^{\text{exp}}>p$ the effect will be opposite.
%In particular, one could visit the search process in its midst ask what is the expected probability of target detection.
At the same time, if the process is already restarted at the optimal rate $r^*$, then $[dp_r/dr]_{r^*}=0$ and small additional correction $\delta r$ to $r^*$ does not change the probability of success $p_{r^*}$ in the leading order approximation.
Therefore, for the optimally restarted process,  $p_{r^*}$ must be equal to  $p_{r^*}^{\text{exp}}=p_{r^*}\langle T_{r^*}^s\rangle/\langle T_{r^*}\rangle$ that immediately leads to  Eq. (\ref{MFPT_identity}).
Interestingly, the match of unconditional and conditional MFPTs is an inherent property of some two-thresholds first-passage processes relevant to kinetics of enzyme reactions \cite{Qian_2006,Ge_2008}, motor proteins dynamics \cite{Kolomeisky_2005}, entropy-production fluctuations \cite{Neri_2016} and decision making \cite{Roldan_2015}. 
From the foregoing considerations it follows that for all these processes the splitting probabilities coincide with the corresponding expected splitting probabilities.

%\textbf{Deterministic restart is the optimal strategy.}
%So far we did not made any assumption regarding the underlying process, but restart was assumed to be Poisson with time independent rate parameter. 
%Now let us abandon the assumption of Poisson restart and return to Eq. (\ref{mean_x}), which is valid for a generic restart mechanism. 
Anticipating that optimization is not an exclusive prerogative of Poisson restart, it is natural to ask how to choose a restart time distribution $P^r(R)$ which provides the maximum probability of success $p_r$ for a given first-passage process.
%Now let us consider a broader optimization problem:  
%What is the optimal restart strategy which provides the maximum probability of success for a given stochastic process?
Recently it was proven that deterministic restart (i.e. $P^r(R)=\delta(R-t)$) always outperforms  stochastic restart strategies in terms of attaining the lowest MFPT \cite{Reuveni_PRL_2017}.
%sharply restarted whenever a time t passes from the previous restart (or start) epoch
Arguments similar to those used in \cite{Reuveni_PRL_2017} allow us to conclude that deterministic restart is also  universally  preferable  when one needs to optimize the splitting probabilities.
%Assume that $t^*$ is the optimal period of deterministic restart maximizing the probability of success which is given  by Eq. (\ref{mean_x}), i.e. $\frac{Pr(T<t^*|\text{success})}{Pr(T<t^*)}\ge\frac{Pr(T<t|\text{success})}{Pr(T<t)}$ for all $0<t<\infty$.
%It can be shown then that the inequality $\frac{Pr(T<t^*|\text{success})}{Pr(T<t^*)}\ge\frac{Pr(T<R|\text{success})}{Pr(T<R)}$ is  valid regardless of the distribution of the stochastic restart time $R$.
%% (see Supplementary Material). 
It can be shown that if there exists such $t^*$ that deterministic restart with restart time distribution  $P^r(R)=\delta(R-t^*)$ brings  the probability of success to a maximum $p_{t^*}$, then the value $p_{t^*}$ cannot be exceeded  by  stochastic restart strategies (see Supplementary Material).

%The strategy implies that the process is restarted whenever a
%time $t$ passes, i.e. .
%Assume that $t^*$ is the optimal period of deterministic restart maximizing the probability of success which is given  by Eq. (\ref{mean_x}), i.e. $\frac{Pr(T<t^*|\text{success})}{Pr(T<t^*)}\ge\frac{Pr(T<t|\text{success})}{Pr(T<t)}$ for all $0<t<\infty$.
%It can be shown then that the inequality $\frac{Pr(T<t^*|\text{success})}{Pr(T<t^*)}\ge\frac{Pr(T<R|\text{success})}{Pr(T<R)}$ is  valid regardless of the distribution of the stochastic restart time $R$. 
%This proves that the deterministic restart strategy is optimal among all possible
%stochastic restart strategies.
%%optimal in the entire space of restart distributions
%Figure show the probability of target detection by mortal Brownian searcher as a function of restart frequency for different restart mechanisms. 
%Equations (\ref{mean_x}) and (\ref{MFPT}) allow us also to compute the splitting probabilities and conditional first-passage time of an arbitrary stochastic process under generic restart mechanism.
%What can be said in general  regardless of the particular details of the process or restart mechanism.
% inescapable

%The  price one has to pay for attaining the higher probability of success via optimal deterministic restart  is the relatively longer average completion time of successful trials.
%Indeed,  
%Instead of  Eq. (\ref{MFPT_identity}) found in the case of optimal stochastic restart,
Equation (\ref{MFPT_identity}) is not longer valid when restart events have non-Poisson statistics. 
%If restart events have non-Poisson statistics, Eq. (\ref{MFPT_identity}) is no longer valid.
Instead, the conditional and  unconditional mean first-passage times of a process undergoing optimally tuned deterministic restart obey the universal inequality constraint
%It is possible to prove, that for the general stochastic process under optimal deterministic restart the equalities in Eq. (\ref{MFPT_identity}), which is valid only for Poisson statistics of restart events, are replaced by the inequalities so that
\begin{equation}
\label{MFPT_enequality}
\langle T_{t^*}^s\rangle\ge \langle T_{t^*}\rangle.
\end{equation}
%which replaces Eq. (\ref{MFPT_identity}) derived under an assumption of Poisson statistics of restart events. 
%It is easy to check that  Eq. (\ref{MFPT_identity}) is no longer valid when a process is restarted in non-Poisson way.   
%It is possible, however, to prove that for the arbitrary process under optimal deterministic restart the chain of inequalities 
%\begin{equation}
%\langle T_{t^*}^F\rangle\le \langle T_{t^*}\rangle\le \langle T_{t^*}^S\rangle
%\end{equation}
% is valid.  
To prove Eq. (\ref{MFPT_enequality}), let us assume that the process, which is being restarted deterministically in an optimal way,  becomes subject to additional Poisson restart with an infinitesimally small rate $\delta r$.
That produces a deferential correction $\delta p$ to the  probability of success $p_{t^*}$ attained by deterministic restart.
  % the applying of additional restart mechanism cannot  increase $p_r$.
Equation (\ref{conditional_MFPT}) allows us to write $\langle T_{t^*}\rangle-\langle T_{t^*}^s\rangle=\delta p/(p_{t^*}\delta r)$.
%Since the value $p_{t^*}$ cannot be exceeded 
Due to dominance of deterministic restart over other restart strategies,
% in terms of probability of success, 
one can be sure that $\delta 
p\le 0$ and, therefore, $\langle T_{t^*}^s\rangle\ge\langle T_{t^*}\rangle$.
Taking into account the identity  $p_r\langle T_{r}^{s}\rangle+(1-p_r)\langle T_{r}^{f}\rangle=\langle T_r\rangle$, one  arrives at the opposite inequality for the MFPT of failed trials: $\langle T_{t^*}^f\rangle\le\langle T_{t^*}\rangle$.
These features of deterministic restart are clearly seen in the right panel of  Fig \ref{pic:MFPT}.

%Equation (\ref{MFPT_enequality}) is also robust to appearance of time penalty for restart events, see SI. 

%Another noteworthy optimal property of the deterministic restart is that it requires the smallest mean number of restart events 
%given a fixed mean restart time.
%Indeed, the mean number of restarts till process completion is given by $\langle N\rangle=\frac{Pr(T> R)}{Pr(T\le R)}=\langle \int_0^R dTP(T)\rangle_R^{-1}-1$, see SI.
%Since function 

\textit{Conclusion.}---
First-passage processes exhibiting stochasticity not only in the timing of their evolution but also in the very result of the evolution are ubiquitous in science.
%First-passage processes exhibiting stochasticity not only in the time required for the completion of their evolution  but also in the very result of the evolution are ubiquitous in science.
%Processes exhibiting stochastic variations not only in the duration of their dynamics but also in the very result of the dynamics are ubiquitous.
%First-passage processes having several possible outcomes are ubiquitous in science.  
%Plurality of possible outcomes may arise from the competition among several different first-passage  phenomena or due to multiple thresholds for one and the same first-passage mechanism.
When the possible outcomes of a process are not equally valuable, the splitting probabilities may come to the fore as a crucial measure of efficiency and reliability \cite{Campos_2015,Roldan_2015,Meerson_2015,Rehbein_2011,Rehbein_2015,Martin-Soomer_2016,Grebenkov_2016_1}. 
% in single-molecule measurements 
%- the measure of relative frequency of different outcomes - may become 
%That results in multiple possible scenarios of process completion which may be not equally desirable when considered from the view of system performance. 
%These quantities are especially useful to quantify the kinetics of competitive
%reactions \cite{Rice_2001,Talkner_2012} and recently has been used to extract information at single molecule level. 
In this Letter we applied a general theoretical approach to describe the effect of restart on the splitting probabilities of a process with exactly two possible completion scenarios.
It is shown that a carefully chosen rate of Poisson restart could maximize (minimize) the probability that the process will complete in the desirable (undesirable) way. 
% . studied an effects of restart on the splitting probabilities of a arbitrary complex  process which has two possible outcomes. 
%The splitting probability are determined by complex interplay of different internal factors.
%In this letter we demonstrated that restart may have profound consequences not only on the first-passage time of a stochastic process, but also on its splitting probabilities.   
% and reveal universal features of the optimally restarted processes.
%The main goal was to reveal the universal aspects of this kind of optimal behaviour.  
 Whenever it is the case, the conditional and unconditional mean completion times are  equal to each other.
We also established the global dominance of deterministic restart in the entire space of restart strategies - further evidence of the great optimization potential of deterministic restart in first-passage problems \cite{Pal_2016,Reuveni_PRL_2017}.
Note that these conclusions are robust to appearance of  a generally distributed random time penalty for restart (see Supplementary Material). 
Thus, our work adds to the collection of universal results in the field of first-passage phenomena \cite{Condamin_2007_2,Benichou_2010,Chupeau_2015,Grebenkov_2016,Reuveni_PRL_2016,Reuveni_PRL_2017}.
% and may find numerous applications.

%Splitting probabilities are also the informative metrics used to characterize the structural and dynamic properties of complex networks \cite{Li_2013, Tavani_2016} and the conformational states of biological macromolecules \cite{Chodera_2011,Guerin_2012,Guerin_2013,Neupane_2015, Manuel_2015,Neupane_2016}.

Of many implications of above results, let us emphasize the issue relevant to chemical kinetics. % is particularly interesting.
The two fundamental problems of chemistry are control over reaction rate \cite{Hanggi_1990} and product selectivity \citep{Rehbein_2011,Rehbein_2015,Martin-Soomer_2016}. % are two basic problems of chemistry. 
As we know thanks to the recent study of enzymatic reactions \cite{Reuveni_2014}, restart of catalytic step can potentially accelerate the rate of product formation.
%It has been shown in the studies of enzymatic reactions that restart of catalytic step provides the possibility to control over rate of product formation. % in chemical reaction. 
%The present work lead to the complementary conclusion that the introduction of restart mechanism may also allow to control over product selectivity in situations when competing pathways of a chemical reaction end up with different products.
The results of the present work lead to the complementary conclusion that when competing pathways of a chemical reaction  end up with different products the introduction of restart mechanism may allow to control over product selectivity. 
This is also relevant to the protein folding reactions in which a single protein molecule can fold in one of distinct native conformations \cite{Solomatin_2010,Marek_2011,Hyeon_2012,Paudel_2014,Hinczewski_2016,Pierse_2017}. 
%One could potentially reduce the degree of resulting conformational heterogeneity by initiating refolding events \cite{Solomatin_2010,Liu_2013,Paudel_2014_1} with carefully adjusted frequency.
%One could potentially optimize the probabilities of getting different conformational states by initiating denaturation events  \cite{Solomatin_2010,Liu_2013,Paudel_2014_1} with carefully adjusted frequency.
One could potentially optimize the probabilities of getting 
different conformational states by initiating the protein refolding that follows the denaturation events \cite{Solomatin_2010,Liu_2013,Paudel_2014_1} with carefully adjusted frequency.
%One could potentially optimize the probabilities of getting different conformational states by tuning the frequency of denaturation events followed by refolding \cite{Solomatin_2010,Liu_2013,Paudel_2014_1}.

%,  i.e. branching ratio, in a chemical reaction. 
%As one of possible applications we 
%We have only been concerned with the MFPT of restarted process. 
%The fluctuations remains unexplored. 
%The approach allows to find how restart modifies the probability densities of the different outcomes. 

%This work does not cover all variety of questions arising when one considers the effect of restart on a first-passage process with plural outcomes. 
This paper covers only some of many interesting questions relating to the effect of restart on the first-passage processes with plural outcomes. 
Particularly, we have only considered the mean conditional first-passage times, thus, leaving aside the issue of fluctuations.
Besides, it would be also interesting to consider the problem of optimization over both the splitting probabilities and the mean first-passage time because in some settings the time required for process completion is not less important than the outcome of the process. 
Finally, note that while looking for the optimal restart strategy we did not take into account possible cost associated with its implementation \cite{Husain_2016}.
The tradeoffs between performance and cost in optimization problems involving first-passage processes under restart represent an important challenge for future studies.

\textbf{Acknowledgments.}  The author acknowledges stimulating discussions with S. Redner and M. Tolpina.

{}

\begin{widetext}

\newpage

\section{Supplementary information}

\subsection{Mortal Brownian searcher in one dimension}

Assume that a mortal searcher undergoes diffusion starting from the initial position $x_0\ge 0$.
The search process ends either with detection of a target at $x=0$ or with searcher' death. 
The probability that the searcher will die in the time interval $[t,t+dt]$ is given by $\alpha dt$ where $\alpha$ is the time-independent mortality rate.  
The probability distribution of the completion time is known to be (see \cite{Meerson_2015})
\begin{equation}
\label{FPTD_searcher}
P(T)=\frac{x_0}{\sqrt{4\pi DT^3}}e^{-\alpha T-x_0^2/4DT}+\alpha e^{-\alpha T}\erf\left(\frac{x_0}{\sqrt{4DT}}\right),
\end{equation}
where $D$ is the diffusion constant. 
The first term in the right hand side of Eq. (\ref{FPTD_searcher}) comes from those realizations of the search process which end with target detection (success), while the second terms is due to mortality (failure).

Now assume that the searcher is subject to stochastic reset to the initial position $x_0$ at the rate $r$.
The evolution of the probability distribution $\rho(x,t)$ of the searcher is governed by
the following equation
\begin{equation}
\partial_t\rho=D\partial_x^2\rho-(\alpha+r)\rho+r\delta(x-x_0)\int_0^\infty dy\rho(y,t),
\end{equation}
supplemented by the initial condition $\rho(x,0)=\delta(x-x_0)$ and the boundary condition $\rho(0,t)=0$.
After the Laplace transform 
\begin{equation}
\tilde \rho_s(x)=\int_0^\infty e^{-st}\rho(x,t)dt,
\end{equation}
we obtain
\begin{equation}
s\tilde\rho_s-\delta(x-x_0)=D\partial_x^2\tilde\rho_s-(r+\alpha)\tilde\rho_s+r\delta(x-x_0)\int_0^\infty dy\tilde\rho_s(y).
\end{equation}
Imposing the zero boundary condition at $x=0$ and $x\to+\infty$ together with the continuity condition at $x=x_0$ we find
\begin{equation}
\tilde\rho_s(x)=\left\{ \begin{array}{ll}
A\sinh \gamma_sx, \ \ \ \ \ \ \ \ \ \ \ \ \ \ \ \ x\le x_0,\\
Ae^{\gamma_s(x_0-x)}\sinh \gamma_sx_0,  \ \ \ x>x_0,
\end{array} \right.
\end{equation} 
where $\gamma_s=\sqrt{(s+r+\alpha)/D}$.
To calculate the unknown coefficient $A$ we should take into account the jump of the derivative $\partial_x \rho_s$ at $x=x_0$.
That gives 
\begin{equation}
A=\frac{\gamma_s}{(s+\alpha)e^{\gamma_sx_0}+r}.
\end{equation}
Then the Laplace transforms of the flux to the target $j(t)=D\partial_x\rho(x,t)$ is given by
\begin{equation}
\tilde j(s)=D[\partial_x \tilde\rho(x)]_{x=0}=\frac{s+r+\alpha}{(s+\alpha)e^{\gamma_sx_0}+r}.
\end{equation}
%\begin{equation}
%j_d(s)=\frac{\alpha e^{\gamma_sx_0}-\alpha}{(s+\alpha)e^{\gamma_sx_0}+r}.
%\end{equation}
The probability that the target is eventually found is
\begin{equation}
p_r=\int_0^\infty j(t)dt=\tilde j(0)=\frac{r+\alpha}{\alpha e^{\sqrt{(r+\alpha)/D}x_0}+r},
\end{equation}
which coincides with Eq. (\ref{p_r}) in the main text.
%The probability that searcher dies is
%\begin{equation}
%p_d=j_d(0)=\frac{\alpha e^{\sqrt{(r+\alpha)/D} x_0}-\alpha}{\alpha e^{\sqrt{(r+\alpha)/D}x_0}+r}.
%\end{equation}
%Note that $p_t+p_d=1$.
Alternatively, one can derive this result from Eq. (\ref{splitting_probability}).
Calculating the Laplace transforms  of the probability densities of successful and failed trials, which are $P^s(T)=\sqrt{x_0^2/4\pi DT^3}e^{-\alpha T-x_0^2/4DT}$ and $P^f(T)=\alpha e^{-\alpha T}\erf(\sqrt{x_0^2/4DT})$ in accordance with Eq. (\ref{FPTD_searcher}),  we find 
\begin{equation}
\tilde P^s(s)=\sqrt{\frac{x_0^2}{4\pi D}}\int_0^\infty  T^{-3/2}\exp\left(-(s+\alpha) T-\frac{x_0^2}{4DT}\right)dT=e^{-\sqrt{(\alpha+s)/D}x_0},
\end{equation}
\begin{equation}
\tilde P^f(s)=\alpha\int_0^\infty e^{-(\alpha+s) T}\erf\left(\sqrt{\frac{x_0^2}{4DT}}\right)dT=\frac{\alpha}{\alpha+s}(1-e^{-\sqrt{(\alpha+s)/D}x_0}).
\end{equation}
The ratio $\tilde P^s(r)/\tilde P(r)$, where $\tilde P(r)=\tilde P^s(r)+\tilde P^f(r)$, is equal to the right hand side of Eq. (\ref{p_r}).

\subsection{Derivation of Eq. (\ref{conditional_MFPT}) }

Let us derive Eq. (\ref{conditional_MFPT})   for a more general situation than the one described in the main text. 
Namely, we assume that each restart event entails a generically distributed random time penalty $T_{\text{on}}$. % which is a random variable having generic probability distribution $P_{\text{on}}(T_{\text{on}})$. 
This complication does not restrict the applicability of our arguments concerning the splitting probabilities so that Eqs. (\ref{eq_x}), (\ref{mean_x}) (\ref{splitting_probability}) from the main text remain unchanged.  
However, the time penalty will definitely affect the completion time $T_r$ and Eq. (\ref{fpt}) now becomes  
%The completion time of the restarted process obeys the renewal law
\begin{equation}
\label{fpt_1}
T_r=T_{{\text{on}}}+I(T\ge R)(R+T_r')+I(T<R)T,
\end{equation}
We take expectation of Eq. (\ref{fpt_1}) to find 
\begin{equation}
\label{MFPT_1}
\langle T_{r}\rangle=\frac{\langle T_{{\text{on}}}\rangle+\langle\min (T,R)\rangle}{\langle I(T<R)\rangle}.
\end{equation}
As it follows from \cite{Reuveni_PRL_2016}, when the restart  time $R$ has exponential distribution $P^r(R)=re^{-rR}$, Eq. (\ref{MFPT_1}) reduces to   
\begin{equation}
\label{MFPT_2}
\langle T_{r}\rangle=\frac{r\langle T_{{\text{on}}}\rangle+1-\tilde P(r)}{r\tilde P(r)}.
\end{equation}
Next inserting Eqs. (\ref{eq_x}) and (\ref{fpt_1}) into the definition of the conditional first-passage time $\langle T_r^s\rangle=\langle xT_r\rangle/\langle x\rangle$ we obtain 
\begin{equation}
\label{MFPT_3}
\langle T_{r}^s\rangle=\frac{\langle T_{{\text{on}}}\rangle+\langle I(T>R)R\rangle}{\langle I(T<R)\rangle}+\frac{\langle I(T<R)y_T T\rangle}{\langle I(T<R)y_T\rangle}.
\end{equation}
For Poisson restart, %the success probability $p_r=\langle x\rangle$ is given by Eq. (\ref{splitting_probability}), while 
the averages in the right hand side of  Eq. (\ref{MFPT_3}) can be transformed in the following way 
\begin{equation}
\langle I(T<R)\rangle=r\int_0^\infty\int_0^R e^{-rR}P(T)dRdT=\int_0^\infty e^{-rR}P(R)dR=\tilde P(r),
\end{equation}
\begin{equation}
\langle I(T<R)y_T\rangle=r\int_0^\infty\int_0^R e^{-rR}P^s(T)dRdT=\int_0^\infty e^{-rR}P^s(R)dR=\tilde P^s(r),
\end{equation}
\begin{eqnarray}
\langle I(T>R)R\rangle=r\int_0^\infty\int_R^\infty R e^{-rR}P(T)dRdT=\int_0^\infty\int_R^\infty e^{-rR}P(T)dRdT-\int_0^\infty R e^{-rR}P(R)dR=\\
=\int_0^\infty\int_R^\infty e^{-rR}P(T)dRdT-\int_0^\infty R e^{-rR}P(R)dR=\frac{1}{r}-\frac{1}{r}\int_0^\infty e^{-rR}P(R)dR +\frac{d}{dr}\int_0^\infty  e^{-rR}P(R)dR=\\
=\frac{1}{r}-\frac{1}{r}\tilde P(r)+\frac{d \tilde P(r)}{dr},
\end{eqnarray}
\begin{eqnarray}
\langle I(T<R)y_T T\rangle=r\int_0^\infty\int_0^R T e^{-rR}P^s(T)dRdT=\int_0^\infty R e^{-rR}P^s(R)dR= -\frac{d}{dr}\int_0^\infty e^{-rR}P^s(R)dR=\\
=-\frac{d \tilde P^s(r)}{dr},
\end{eqnarray}
where we have used repeated integration by parts.
Substituting these expressions back into Eq. (\ref{MFPT_3}) yields
\begin{equation}
\label{MFPT_4}
\langle T_{r}^s\rangle=\frac{\langle T_{{\text{on}}}\rangle}{\tilde P(r)}+\frac{1-\tilde P(r)}{r\tilde P(r)}+\frac{1}{\tilde P(r)}\frac{d \tilde P(r)}{dr} -\frac{1}{\tilde P^s(r)}\frac{d \tilde P^s(r)}{dr}=\frac{r\langle T_{{\text{on}}}\rangle+1-\tilde P(r)}{r\tilde P(r)}-\frac{d }{dr}\ln\frac{\tilde P^s(r)}{\tilde P(r)}.
\end{equation}
Recalling Eqs. (\ref{splitting_probability}) and (\ref{MFPT_2}) we see that Eq. (\ref{MFPT_4}) can be rewritten in the form $\langle T_{r}^s\rangle=\langle T_{r}\rangle-d\ln p_r/dr$ which coincides with Eq. (\ref{conditional_MFPT}) from the main text .

\subsection{Expected probability of success $p^{\text{exp}}$}

Consider a first-passage process which starts
at time zero and repeats itself over and over indefinitely. 
Each of the independent trials can end either with success or failure. 
Let $\{t_i\}_{i=1}^\infty=t_1,t_2,t_3,\dots$ be the ordered sequence of  time instants when completion events occur ($\forall$ $i$: $t_i>0$ ).
Then, the completion time of the $i$th trial is given by $T_i=t_{i}-t_{i-1}$. 
Also, let $\{x_i\}_{i=1}^\infty=x_1,x_2,x_3,\dots$ be a string of binary variables encoding the results of trials.
Namely, $x_i=1$ if the $i$th trial is successful, and $x_i=0$ in the case of failure. 

The question that we want to address is the following.
If one starts to observe the process at a random point in time $t$, what is the expected probability $p^{\text{exp}}$ of successful completion of the ongoing trial?
Given sequences $\{t_i\}_{i=1}^\infty$ and $\{x_i\}_{i=1}^\infty$, we can construct a piece-wise process $X(t)=x_{N(t)+1}$, where $N(t)$ is the number on completion events up to time $t$. 
It is easy to see that $X(t)$ equals to unity if the next outcome is success and  zero otherwise.
The expected probability of success $p^{\text{exp}}$ is simply the time average of $X(t)$:
\begin{eqnarray}
p^{\text{exp}}=\lim_{t\to\infty}\frac{1}{t}\int_0^t X(t)dt=\lim_{t\to\infty}\frac{1}{t}\left(\sum_{i=1}^{N(t)}x_iT_i+x_{N(t)+1}(t-t_{N(t)})\right)=\lim_{t\to\infty}\frac{1}{t}\sum_{i=1}^{N(t)}x_iT_i. %=
%=\lim_{t\to\infty}\frac{N(t)}{t}\lim_{t\to\infty}\frac{\sum_{i=1}^{N(t)}x_iT_i}{N(t)}
\end{eqnarray}
%Only successful trials contribute to the sum $\sum_{i=1}^{N(t)}x_iT_i$.
The law of large numbers  tells us that
\begin{equation}
\lim_{t\to\infty}\frac{\sum_{i=1}^{N(t)}x_iT_i}{N(t)}=\langle xT\rangle,
\end{equation}
\begin{equation}
\lim_{t\to\infty}\frac{N(t)}{t}=\frac{1}{\langle T\rangle}.
\end{equation}
Therefore $p^{\text{exp}}=\langle xT\rangle/\langle T\rangle$.
Finally, we can rewrite this as $p^{\text{exp}}=p\langle T^s\rangle/\langle T\rangle$, where  $p=\langle x\rangle$ is the probability of success in a single trial and   $\langle T^s\rangle=\langle xT\rangle/\langle x\rangle$ represents the mean completion time of successful trials.

\subsection{Deterministic restart}

Deterministic restart strategy implies that the process is restarted whenever a
time $t$ passes.
In this case $P^r(R)=\delta(R-t)$.
Assume that $t^*$ is the optimal period of deterministic restart maximizing the probability of success which is given  by Eq. (\ref{mean_x}) in the main text, i.e. 
\begin{equation}
\label{determ_1}
\frac{\langle y_TI(T<t^*)\rangle_T}{\langle I(T<t^*)\rangle_T}\ge \frac{\langle y_TI(T<t)\rangle_T}{\langle I(T<t)\rangle_T}
\end{equation}
 for all $0<t<\infty$.
Let us  multiply both sides of Eq. (\ref{determ_1}) by $\frac{\langle I(T<R)|R\rangle_T}{\langle I(T<R)\rangle_{T,R}}$
\begin{equation}
\label{determ_3}
\frac{\langle I(T<R)|R\rangle_T}{\langle I(T<R)\rangle_{T,R}}\frac{\langle y_TI(T<t^*)\rangle_T}{\langle I(T<t^*)\rangle_T}\ge \frac{\langle I(T<R)|R\rangle_T}{\langle I(T<R)\rangle_{T,R}}\frac{\langle y_TI(T<t)\rangle_T}{\langle I(T<t)\rangle_T}
\end{equation}
Next we replace $t$ by a  generally distributed  random time $R$ and average over statistics of $R$ 
\begin{equation}
\label{determ_4}
\left\langle \frac{\langle I(T<R)|R\rangle_T}{\langle I(T<R)\rangle_{T,R}}\frac{\langle y_TI(T<t^*)\rangle_T}{\langle I(T<t^*)\rangle_T}\right\rangle_R\ge \left\langle  \frac{\langle I(T<R)|R\rangle_T}{\langle I(T<R)\rangle_{T,R}}\frac{\langle y_TI(T<R)\rangle_T}{\langle I(T<R)\rangle_T}\right\rangle_R
\end{equation}
Applying the law of total expectation we find 
\begin{equation}
\label{determ_5}
\left\langle\frac{\langle I(T<R)|R\rangle_T}{\langle I(T<R)\rangle_{T,R}}\right\rangle_R=1
\end{equation} 
and
 \begin{equation}
 \label{determ_2}
%=\frac{\langle\langle y_TI(T<t)|R\rangle_T\rangle_R}{\langle %I(T<R)|R\rangle_{T,R}}
\left\langle  \frac{\langle I(T<R)|R\rangle_T}{\langle I(T<R)\rangle_{T,R}} \frac{\langle y_TI(T<t)|R\rangle_T}{\langle I(T<R)\rangle_T}\right\rangle_R=\frac{\langle y_TI(T<R)\rangle_{T,R}}{\langle I(T<R)\rangle_{T,R}}.
\end{equation}
Utilizing  Eqs. (\ref{determ_5}) and (\ref{determ_2}) we obtains from Eq. (\ref{determ_4}) the following inequality
\begin{equation}
\frac{\langle y_TI(T<t^*)\rangle_T}{\langle I(T<t^*)\rangle_T}\ge \frac{\langle y_TI(T<R)\rangle_{T,R}}{\langle I(T<R)\rangle_{T,R}}.
\end{equation}
%which is valid for any probability distribution of random variable $R$.
On the left we see the success probability $p_{t^*}$ attained by optimal deterministic restart, while the right hand side  represents the success probability $p_r$ for a process restarted at a generally distributed random time $R$.
This proves that deterministic restart is optimal  among all possible stochastic restart strategies.

\end{widetext}


\begin{thebibliography}{99}
\bibitem{note_1} Strictly speaking, we imply not only a resetting of coordinate, but also restart of the mortality mechanism. 
This remark is not essential in this particular example where the searcher  has the constant mortality rate $\alpha$ which does not depend on the time elapsed since the previous restart (or start) event.
\end{thebibliography}
\end{document}